\documentclass[12pt]{article}
\usepackage{hyperref}

\newcommand{\be}{\begin{equation}}
\newcommand{\ee}{\end{equation}}
\newcommand{\ba}{\begin{array}}
\newcommand{\ea}{\end{array}}
\newcommand{\bqa}{\begin{eqnarray}}
\newcommand{\eqa}{\end{eqnarray}}
\newcommand{\cO}{{\cal O}}
\newcommand{\mL}{\mathcal{L}}

\newcommand{\mM}{\mathcal{M}}

\newcommand{\Frac}[2]{\frac{\displaystyle #1}{\displaystyle #2}}
\newcommand\lsim{\mathrel{\rlap{\lower4pt\hbox{\hskip1pt$\sim$}}
    \raise1pt\hbox{$<$}}}
\newcommand\gsim{\mathrel{\rlap{\lower4pt\hbox{\hskip1pt$\sim$}}
    \raise1pt\hbox{$>$}}}

\begin{document}

\title{ \bf \boldmath Partial waves and  large $N_C$ resonance sum rules}

\author{
Z.~H.~Guo, J.~J.~Sanz Cillero,  H.~Q.~Zheng
\vspace{0.5cm}\\
Department of Physics, Peking University,  \\
Beijing 100871, P.~R.~China }

\date{\today}
\maketitle

\begin{abstract}
{  Using $1/N_C$ expansion and dispersion theory techniques, without
relying on  any explicit resonance lagrangian, we  generalize the
KSRF relation by including the scalar meson effects, at leading
order of chiral expansion. Two sum rules for the low energy
constants $L_2$, $L_3$ and a new relation between resonance
couplings are also derived. A rather detailed examination to the new
relation is also given. We also discussed the $N_c$ properties  of
partial wave amplitudes and the broad $\sigma$ resonance.
 }
\end{abstract}

\section{Introduction}

Low energy effective field theories (EFT) are useful tools in modern
particle physics~\cite{EFT}. The EFT lagrangian can be obtained
through the integration of the heavy degrees of freedom of the whole
theory. The more interesting and difficult problem is how to
understand ``high energy physics" from the low energy theory. In
hadron physics, the low energy effective theory is chiral
perturbation theory ($\chi$PT) whose degrees of freedom are just the
light pseudo--Goldstone bosons from the spontaneous chiral symmetry
breaking~\cite{chptweinberg,chptoneloop,chptms}. A former paper was
devoted to the study of the inverse problem in hadron physics using
techniques from $S$--matrix theory, low energy effective theory and
$1/N_C$ expansion and it was demonstrated that resonances with
$M\,,\Gamma\sim O(N_C^0)$~\cite{xiao05} could not exist. However,
the crossed channel resonance exchange contribution to the left-hand
cut were not considered in that paper. The present work performs a
large--$N_C$ calculation  of the $\pi\pi$ scattering including
right- and left-hand cut contributions. The analysis is taken up to
next-to-leading order in the chiral expansion. This yields a
consistent set of relations between the chiral couplings related to
$\pi\pi$--scattering and the resonance parameters.

The partial wave amplitudes are extracted in Section 2 through
dispersive relations. We perform a low-energy matching to $\chi$PT
in section 3. A generalized KSRF relation is extracted together with
predictions for the low energy constants (LECs) $L_2$ and $L_3$.
Section 4 studies the consistency of different phenomenological
lagrangians under the generalized KSRF constraint. The influence of
a broad sigma meson, generated through $K$--matrix unitarization of
the current algebra amplitude, is analyzed in Section 5. The results
are discussed and summarized in Section 6.

\section{Dispersive calculation of the $S$--matrix}

The $S$--matrix describing the partial wave elastic
$\pi\pi$--scattering  accepts the general factorization~\cite{piK}
 \be\label{param}
S \, = \, S^{\rm cut}\, \cdot\, \prod_{\rm R} S^{\rm R} \,  ,
 \ee
 where $S^{\rm R}$ are
the simplest $S$--matrices characterizing  isolated singularities on
the second Riemann-sheet that are solutions of the generalized
single-channel unitarity relations~\cite{HXZ02}. It is noticed that
the Eq.~(\ref{param}) is formally rigorous and can be obtained under
the same condition from which the standard partial wave dispersion
is derived. I.e.,
  the so called maximal
 analyticity assumption or Mandelstam representation.

\subsection{Contribution from the $s$--channel poles}
\subsubsection{Resonances in the $s$--channel}

The part of the $S$--matrix that contains the pole singularities
related to second sheet resonances is given by
\begin{equation}
\prod_{\rm R} S^{\rm R}(s)\, = \, \prod_{\rm R}\left(\, 1 \, + \, 2\, i
\,\rho(s) \, T^{\rm sR}(s)\right)\, ,
\end{equation}
with
\begin{equation}
\label{resp} T^{\rm sR}(s)=\frac{s\, G_{\rm R}[z_0]  }{ M_{\rm
R}^2[z_0]-s-i\rho(s)s\, G_{\rm R}[z_0]}\ ,
\end{equation}
where $M_{\rm R}^2[z_0]$ and $G_{\rm R}[z_0]$ are related to the
pole position $z_0\equiv \left(M+\frac{i}{2}\Gamma\right)^2$ of the
resonance R~\cite{piK},
 \bqa\label{MandGamma}
{M^2_{\rm R}}[z_0] &=& \mathrm{Re}[z_0] + \frac{\mathrm{Im}[z_0]\,
     \mathrm{Im}[z_0\,\rho (z_0)]}{\mathrm{Re}[
     z_0\,\rho (z_0)]}\ ,\nonumber \\%
     G_{\rm R}[z_0] &=& \frac{\mathrm{Im}[z_0]}{\mathrm{Re}[z_0\,\rho (z_0)]}\ .
\eqa
The $S$--matrix phase-space factor is defined as
$\rho(s)=\sqrt{1-4 m_\pi^2/s}$, such that for $s>4 m_\pi^2$  one has
the prescription $\rho(s\pm i\epsilon)=\pm |\rho(s)|$. In the paper,
we will refer to $\rho(s)$ as $\rho(s+i\epsilon)$. Notice that real
analyticity requires the existence of a companion pole at $z_0^*$.

When discussing large $N_C$ dynamics, it is not clear whether, in
addition to the narrow width states lying near the physical region,
there are any other $S$ matrix poles with odd behavior.
Nevertheless, the quantity ${G_{\rm R}[z_0]/(M_{\rm R }^2[z_0]-
4m_\pi^2) }$ is always positive definite for any location of the
pole $z_0$ in the complex $s$--plane~\cite{xiao05}. Because of this,
there can be no $S$ matrix poles located on the $s$--plane when
$N_C\to\infty$, except on the real axis or at
infinity~\cite{xiao05}. In most of this paper, we assume that all
$S$--matrix poles indeed move to the real axis when $N_C\to\infty$.
Only in section~\ref{scalarphys.} we will pay some attention to the
possibility that there exists a pole moving to infinity.

The $s$--channel second sheet resonance contribution to the
$T$--matrix is,
\begin{equation}\label{ReTs} T^{\rm sR}(s)\, =\,  G_{\rm R}[z_0]\frac{s}{M_{\rm R}^2[z_0]-s} \, \, \,+\, \,
\mathcal{O}\left(\frac{1}{N_C^2}\right) \,,
\end{equation}
with the resonance parameters given in the large--$N_C$ limit by
\begin{eqnarray}\label{ReTs'}
&&M_{\rm R}^2[z_0]\, = \,
 M_{\rm R}^2\,, \nonumber \\
&&G_{\rm R}[z_0]\, =\, \frac{1}{\rho(M_{\rm R}^2)}\,\,
\frac{\Gamma_{\rm R}}{M_{\rm R} } \ ,
\end{eqnarray}
where $M_{\rm R}$ and $\Gamma_{\rm R}$ are defined as the large
$N_C$ limit of the $z_0$ pole parameters $M$ and $\Gamma$,
respectively.

Eq.~(\ref{resp}) would be modified in the case of resonances lying
beyond the elastic region on higher Riemann sheets. However,
Eqs.~(\ref{ReTs})~and~(\ref{ReTs'}) are still valid   in the
large--$N_C$ limit if one replaces the width $\Gamma_{\rm R}$ by the
partial decay width $\Gamma_{R\to\pi\pi}$.

The imaginary part of $T^{\rm sR}$ in
Eq.~(\ref{ReTs}) shows the standard narrow-width expression
\begin{equation} \label{eq.imtr}
\mbox{Im}T^{\rm sR}(s)\, =\, \pi \,  \frac{M_{\rm R}\,
\Gamma_{\rm R}  }{\rho(M_{\rm R}^2)}\, \, \delta(s-M_{\rm R}^2)\,.
\end{equation}
This expression can be directly extracted from the imaginary part of
Eq.~(\ref{resp}) in the limit $G_{\rm R}[z_0]\to 0$.
Eq.~(\ref{ReTs}) is recovered back through a once-subtracted
$T$--matrix dispersion relation.

The expansion of $\displaystyle{\prod_{\rm R}} S^{\rm R}$ in $1/N_C$
is given at the first non-trivial order by
\begin{equation}\label{sumTsR}
\prod_{\rm R} S^{\rm R}(s)\, =\, 1 \, +\, 2\, i \, \rho(s)\, \sum_R T^{\rm sR}(s) \,
\,\, +\, \, \mathcal{O}\left(\frac{1}{N_C^2}\right)\, .
\end{equation}
It is worth noticing that we start our discussions from an $S$
matrix theory point of view: The width has a non-perturbative
definition and is related to the imaginary part of the pole
position. This is very important since it enables us to investigate
general properties of resonances without recurring to perturbative
calculations of the width. As it will be seen later the resonance sum
rules derived and investigated in this paper are obtained without
making use of resonance chiral lagrangians of any kind. Only when we
apply our relations in lagrangian models,  the latter will be needed.

\subsubsection{Virtual pole in the $IJ=20$ channel}

Contrary to the $IJ=00$ and $IJ=11$ channels,  the $IJ=20$
$S$--matrix contains a virtual pole hidden on the second Riemann
sheet at $s_{v}^{(20)}$, related to a $S$--matrix zero in the first
Riemann sheet~\cite{ang}. The pole position is estimated in the
large--$N_C$ limit  from the $\chi$PT $S$--matrix, $S^{\chi
PT}(s)_{(20)}=1+2 i \rho(s) T^{\chi PT}(s)_{(20)}$:
\begin{eqnarray}
s_v^{(20)} \, &=& \, 16 \, m_\pi^2\, T^{\chi PT}(0)^2\, \, +\, \, \,
\cO(m_\pi^{10}) \nonumber
\\
&= &\, \,
\frac{m_\pi^6}{16\pi^2f^4}+\frac{m_\pi^8}{3\pi^2f^6}(10L_2+2L_3-3L_5
+6L_8) \,\,\, +\,\, \mathcal{O}(m_\pi^{10})\, ,
\end{eqnarray}
where $s_v^{(20)}$ is $\cO(m_\pi^6)$ in the chiral expansion. The
contribution of a virtual $S$--matrix pole can be parameterized as
\begin{equation}
S^{sv}(s)_{(20)}\, =\, 1\, +\,2 \, i\, \rho(s)\, T^{\rm sv}(s)_{(20)}\,
\end{equation}
with the $T$-matrix,
\begin{equation}
T^{\rm sv}(s)_{(20)}\, =\,  \frac{a_v^{(20)}}{1\, -\, i\, \rho(s) \,
a_v^{(20)}}\, .
\end{equation}
The scattering length $a_v^{(20)}$ is related to the virtual pole
position through
\begin{eqnarray}
a_v^{(20)}\, &=& \sqrt{\frac{s_v^{(20)}}{4 m_\pi^2 -s_v^{(20)}}}\, =
\, 2\, T^{\chi PT}(0)_{(20)}\, \, +\, \,\, \cO(m_\pi^6)\, \,
\nonumber
\\
&=&\, \frac{m_\pi^2}{8\pi f^2}+\frac{m_\pi^4}{3\pi f^4}(10L_2+2L_3-3
L_5+6L_8) \, \, \, +\,\, \mathcal{O}(m_\pi^6)\, .
\end{eqnarray}
Hence, at leading order in $1/N_C$, the contribution to the $IJ=20$
$T$--matrix from the virtual pole is
\begin{equation}
T^{\rm sv}(s)_{(20)}\,=\, a_v^{(20)} \, \, \,
+\,\,\mathcal{O}\left(\frac{1}{N_C^2}\right)\,\,\,  =\, \, 2 \,
T^{\chi PT}(0)_{(20)}\, \, \,
+\,\,\mathcal{O}\left(\frac{1}{N_C^2},m_\pi^6\right)\, .
\end{equation}

\subsection{Contribution from the $t$--channel resonance exchange}

The contribution $S^{\rm cut}$ only contains cuts.  It can be
parameterized in the form~\cite{zhou}, \bqa\label{fS'} f(s)\,
&\equiv &\, \frac{1}{2 i\rho(s)} \, \ln{S^{\rm cut}}(s)\,
 \eqa
and $f(s)$ satisfies the following once subtracted dispersion
relation
 \bqa\label{fLR} f(s)=f_L(s)+f_R(s)&\equiv&\frac{s}{\pi}\int_{L}\frac{{\rm
 Im}_Lf(s')}{s'(s'-s)}ds'+\frac{s}{\pi}\int_{R'}\frac{{\rm
 Im}_Rf(s')}{s'(s'-s)}ds'
\ , \label{fS} \eqa
 where $L$ denotes the left-hand cuts and $R'$ denotes the inelastic cuts
 beyond the $\pi\pi$ elastic one. In the large $N_C$ limit this reduces to a
 left-hand cut contribution from the $t$--channel resonance exchange if higher
resonance multiplets are neglected in the $s$--channel .

Naively, one would expect the two-particle left-hand cuts to be
subleading in $1/N_C$.  However,
$\mathrm{Im}_Lf(s)$
 contains a
kinematical singularity at $s=0$.  As the dispersive left-hand
$\pi\pi$ cut runs in the range $(-\infty,0]$, one gets the
contribution~\cite{xiao05}
\begin{equation}\label{pipitoBG}
f(s)_{L,\,\pi\pi}\, =\, - \left|T(0)\right| \, + \,
\mathcal{O}(1/N_C^2)\, ,
\end{equation}
with $T(0)$ the value of the physical $T$--matrix at $s=0$. The
discontinuity of $f(s)$ for the left-hand cut due to the
$t$--channel resonance exchange obeys the relation
 \bqa\label{IMLR}
\mathrm{Im}_L f(s)&=&-\frac{1}{2\rho(s)}\ln|S^{\rm cut}(s)| =-\frac{1}{2\rho(s)}\ln|S(s)|\nonumber\\
&&
 = \mathrm{Im}_L T \, +\,
\mathcal{O}\left(\frac{1}{N_C^2}\right)\, ,
\eqa%
where $\ln{|S(s)|}=\frac{1}{2}\ln{\left[
1-4\rho(s)\mathrm{Im}_{L}T(s)+4\rho^2(s)|T(s)|^2 \right]}$  has been
expanded using $T(s)=\mathcal{O}\left(\frac{1}{N_C}\right)$.
 Since the
cut due to crossed channel resonance exchanges does not contain the
singular point $s=0$, the expansion of the logarithm in $1/N_C$ can
be safely performed. By means of Eqs.(\ref{fS}-\ref{IMLR}), one
finds the left-hand cut contribution to be given by
\begin{equation}\label{pipitoBG'}
f_L(s)\, =\, - \, \left|T(0)\right|\, +\, \sum_R T^{\rm tR}(s)\, \, \,
+\, \, \,\mathcal{O}\left(\frac{1}{N_C^2}\right)\, ,
\end{equation}
with the $t$--channel resonance exchange contribution,
\begin{equation} \label{eq.TtR}
T^{\rm tR}(s)\, =\,\frac{s}{\pi}\int_{-\infty}^{-M_{\rm R}^2+4m_\pi^2}
\frac{{\rm
 Im}T^{\rm tR}(s')}{s'(s'-s)}ds'  \, .
\end{equation}

According to the convention provided by Ref.~\cite{piK}, the
left-hand cut, or the background contribution to the scattering
phase shift is,
 \be \delta_{BG}=\rho(s)f_L(s)\ .
  \ee
\hspace*{0cm} From Eq.~(\ref{pipitoBG}), at large--$N_C$,
there is always a negative contribution $-|T(0)|$ to the scattering lengths.
On the other hand, the contribution from the crossed channel large--$N_C$
resonances varies in different channels.  This will
be further discussed in section~\ref{stl-e-e}.

Crossing symmetry relates the right to the left-hand cut through the
expression~\cite{martin},
\begin{eqnarray}
 \label{imtl} \mathrm{Im_L}T^{I}_{J}(s) &=&
\frac{1+(-1)^{I+J}}{s-4m_{\pi}^2}
\sum_{J'}\sum_{I'}(2J'+1)C^{st}_{II'} \\&& \times \, \,
\int_{4m_{\pi}^2}^{4m_{\pi}^2-s}dt \, \,
P_J(1+\frac{2t}{s-4m_{\pi}^2}) \, P_{J'}(1+\frac{2s}{t-4m_{\pi}^2})
\, \mathrm{Im_R}T^{I'}_{J'}(t) \, , \nonumber
\end{eqnarray}
with $P_n(x)$ the Legendre polynomials. In general, this
representation is only valid for the range  $-32
m_\pi^2<s<0$  if the Mandelstam representation is assumed~\cite{martin}.
Nevertheless, in the large--$N_C$ limit, Eq.~(\ref{imtl}) actually work for any energy
since the double spectral function vanishes at this order of the $1/N_C$ expansion.
The crossing matrix is given by~\cite{martin}
\begin{equation}
C^{(st)}_{II'}\, =\, \left(\begin{array}{rrr} 1/3 & 1 & 5/3
 \\ 1/3 & 1/2 & -5/6
 \\ 1/3 & -1/2 & 1/6
\end{array}\right)\, .
\end{equation}

Substituting the narrow-width right-hand cut expression from
Eq.(\ref{eq.imtr}), one gets the contribution from the $t$--channel
exchange of a resonance $R$ with $I'J'$ quantum numbers:
\begin{eqnarray}
\mathrm{Im}T^{\rm tR}(s)^{I}_{J}  &=& \,\, \theta(-s-M_{\rm R}^2+4m_\pi^2)\,
\,\times \, \frac{1+(-1)^{I+J}}{s-4m_{\pi}^2} (2J'+1)C^{st}_{II'}
\\
&& \times \, \, P_J(1+\frac{2 M_{\rm R}^2}{s-4m_{\pi}^2}) \,
P_{J'}(1+\frac{2s}{M_{\rm R}^2-4m_{\pi}^2}) \,\, \frac{\pi\, M_{\rm R}\,
\Gamma_{\rm R}}{\rho(M_{\rm R}^2)}\, . \nonumber
\end{eqnarray}
In our analysis, only vector and scalar resonances are considered.
Their contributions to the different channels are obtained through Eq.~(\ref{eq.TtR}):
\begin{enumerate}

\item{$IJ=11$ channel}

\begin{eqnarray}
T^{\rm tS}(s)=&&\frac{2M_{\rm S}\Gamma_{\rm S}}{3\rho(M_{\rm S}^2)}   \left[\frac{-s}{2m_\pi^2(s-4m_\pi^2)}+
\frac{2m_\pi^2-M_{\rm S}^2}{8m_\pi^4}\ln{\frac{M_{\rm S}^2-4m_\pi^2}{M_{\rm S}^2}}
\right. \nonumber \\
&&  \qquad \left. +
\frac{s+2M_{\rm S}^2-4m_\pi^2}{(s-4m_\pi^2)^2}\ln{\frac{s+M_{\rm S}^2-4m_\pi^2}{M_{\rm S}^2}}   \right]\,  ,
\end{eqnarray}

\bqa
T^{\rm tV}(s)=&&\frac{3M_{\rm V}\Gamma_{\rm V}}{\rho(M_{\rm V}^2)}
\left[\frac{-s(M_{\rm V}^2+4m_\pi^2)}{2m_\pi^2(s-4m_\pi^2)(M_{\rm V}^2-4m_\pi^2)}
\right. \nonumber
\\ &&
\qquad +\frac{8m_\pi^4-6m_\pi^2M_{\rm V}^2+M_{\rm V}^4}{8m_\pi^4(4m_\pi^2-M_{\rm V}^2)} \ln{\frac{M_{\rm V}^2-4m_\pi^2}{M_{\rm V}^2}}
\nonumber
\\ &&
+\frac{16m_\pi^4-12m_\pi^2s-12m_\pi^2M_{\rm V}^2+5M_{\rm V}^2s+2M_{\rm V}^4+2s^2}{(s-4m_\pi^2)^2(M_{\rm V}^2-4m_\pi^2)}
\nonumber \\ &&
\qquad \qquad \left. \times \,
\ln{\frac{s+M_{\rm V}^2-4m_\pi^2}{M_{\rm V}^2}} \right]\,  ,
\eqa

\item{$IJ=00$ channel}
\bqa
T^{\rm tS}(s)=&&\frac{2M_{\rm S}\Gamma_{\rm S}}{3\rho(M_{\rm S}^2)}
\left[
\frac{1}{4m_\pi^2}\ln{\frac{M_{\rm S}^2-4m_\pi^2}{M_{\rm S}^2}} +
\frac{1}{s-4m_\pi^2}\ln{\frac{s+M_{\rm S}^2-4m_\pi^2}{M_{\rm S}^2}}\right ]  ,\nonumber \\
\eqa

\bqa
T^{\rm tV}(s)=&&\frac{6M_{\rm V}\Gamma_{\rm V}}{\rho(M_{\rm V}^2)}
\left[
\frac{1}{4m_\pi^2}\ln{\frac{M_{\rm V}^2-4m_\pi^2}{M_{\rm V}^2}}
+\frac{2s+M_{\rm V}^2-4m_\pi^2}{(s-4m_\pi^2)(M_{\rm V}^2-4m_\pi^2)}
\right. \nonumber
\\ && \qquad \qquad\qquad \qquad
\left.  \times \,
\ln{\frac{s+M_{\rm V}^2-4m_\pi^2}{M_{\rm V}^2}}   \right] \, ,
\eqa

\item{$IJ=20$ channel}
\bqa
T^{\rm tS}(s)=&&\frac{2M_{\rm S}\Gamma_{\rm S}}{3\rho(M_{\rm S}^2)}
\left[
\frac{1}{4m_\pi^2}\ln{\frac{M_{\rm S}^2-4m_\pi^2}{M_{\rm S}^2}} +
\frac{1}{s-4m_\pi^2}\ln{\frac{s+M_{\rm S}^2-4m_\pi^2}{M_{\rm S}^2}}  \right] , \nonumber \\
\eqa

\bqa
T^{\rm tV}(s)=&&\frac{-3M_{\rm V}\Gamma_{\rm V}}{\rho(M_{\rm V}^2)}
\left[
\frac{1}{4m_\pi^2}\ln{\frac{M_{\rm V}^2-4m_\pi^2}{M_{\rm V}^2}}
+\frac{2s+M_{\rm V}^2-4m_\pi^2}{(s-4m_\pi^2)(M_{\rm V}^2-4m_\pi^2)}
\right.
\nonumber \\  && \qquad\qquad \qquad \qquad  \left.
\times \, \ln{\frac{s+M_{\rm V}^2-4m_\pi^2}{M_{\rm V}^2}}   \right] \, ,
\eqa

\end{enumerate}
where $T^{\rm tS}$ and $T^{\rm tV}$ denote the contributions from
scalar and vector resonances, respectively.

\subsection{Summation of right- and left-hand cuts}

Putting all the different contributions at leading order in $1/N_C$ together one
gets
\begin{equation}
S(s)\, =\, S^{\rm cut}(s) \, \cdot\, \prod_{\rm R} S^{\rm R}(s) \, =\, 1 \, + \, 2\,
i\, \rho(s) \, T(s)_{N_C\to\infty}\,\, \,  + \,\,\,
\mathcal{O}\left(\frac{1}{N_C^2}\right) \, ,
\end{equation}
with the large--$N_C$ $T$--matrix given by
\begin{equation} \label{eq.sumT}
T(s)_{N_C\to\infty}\, =\, \sum_{\rm R} T^{\rm sR}(s) \,+ \, T^{\rm sv}(s) \,- \,
\left|T(0)\right|\,  +  \, \sum_{\rm R} T^{\rm tR}(s)\, .
\end{equation}
This expression can be simplified taking into account that, in the
channels $IJ=00$ and $IJ=11$, there is no virtual pole
($T^{\rm sv}(s)=0$)  and $\chi$PT tells us that $|T(0)|=-T(0)$.  In the
$IJ=20$ case, $\chi$PT dictates $|T(0)|=T(0)$ and the virtual
pole   contribution $T^{\rm sv}(s)=2 \, T(0)$. Thus,
Eq.~(\ref{eq.sumT}) can be rewritten in the way
\begin{equation}\label{eq.solution'}
T(s)_{N_C\to\infty}\, =\,  T(0)\, + \, \sum_R T^{\rm tR}(s) \, + \,
\sum_R T^{\rm sR}(s)\,.
\end{equation}
An alternative way to reach this relation is through the $T$--matrix
dispersive relation
\begin{equation}\label{eq.solution}
T(s)\,=\, T(0)\, + \frac{s}\pi\int_{-\infty}^0\frac{ds' \,
\mbox{Im}T(s')}{s'(s'-s)}\,  +\, \frac{s}{\pi}\int_{4
m_\pi^2}^\infty  \frac{ds' \, \mbox{Im}T(s')}{s'(s'-s)}\,.
\end{equation}
The above derivation demonstrates that the dispersive
parametrization in Eq.~(\ref{param})~\cite{xiao05} is equivalent to
a $T$--matrix partial wave dispersion relation under narrow width
approximation. The PKU parametrization form is, in this sense,
simply a combination of partial wave dispersion relation and single
channel unitarity.

\section{Low-energy matching}

\subsection{Low-energy expansion of the $s$-- and $t$--channel
resonance contributions}\label{stl-e-e}

We now intend to perform a matching of our dispersive expression in
Eq.(\ref{eq.solution'}) to low-energy QCD, provided by Chiral
Perturbation Theory ($\chi$PT)~\cite{chptoneloop}. 
 Hence, we perform a threshold expansion in the form
\begin{equation}
A(s)=\sum_{n=0}^\infty a_{2n}\,
\left(\frac{s-4m_\pi^2}{m_\pi^{2}}\right)^n\
 .
\end{equation}
The constants $a_n$ are functions of $m_\pi^2$ and can be also
chiral expanded in the form
\begin{equation}
a_{2n}\, =\, \sum_{k=0}^\infty \, a_{2n,2k}\, (m_\pi^2)^k\, .
\end{equation}
To match $\chi$PT up to a given order
$\mathcal{O}\left(p^{2\ell}\right)$ means to match the corresponding
coefficients $a_{2n,2k}$ for $n=0...\ell$, $k\leq \ell-n$.

Taking the  result from Eq.(\ref{eq.solution'}) to low energies and
matching $\chi$PT leads to the relation
\begin{equation} \label{eq.Tmatch}
\begin{array}{l}
t_0^{\chi PT}  - T^{\chi PT}(0)\, +\, t_2^{\chi PT}\,
\left(\frac{s-4m_\pi^2}{m_\pi^2}\right)\, + \, t_4^{\chi PT}\,
\left(\frac{s-4m_\pi^2}{m_\pi^2}\right)^2 + ...
\\
\\
= \, \,[t_0^{s}+t_0^t]  \, +\, [t_2^s+t_2^t]\,
\left(\frac{s-4m_\pi^2}{m_\pi^2}\right)\, + \, [t_4^s+t_4^t]\,
\left(\frac{s-4m_\pi^2}{m_\pi^2}\right)^2 + ...
\end{array}
\end{equation}
The scattering amplitude $T(s)$ on the left-hand side of
Eq.(\ref{eq.solution}) and $T(0)$  have been
substituted by their value in $\chi$PT. The matching is performed in
this work up to $\mathcal{O}(p^4)$. The expansion of the right-hand
cut contribution $\sum_R T^{\rm sR}(s)=\sum_n t^s_{2n} \left(\frac{s-4
m_\pi^2}{m_\pi^2}\right)^n$ is provided by the coefficients
\begin{eqnarray}\label{compare2}
t^s_{0}&=& \sum_R \frac{1}{\rho(M_{\rm R}^2)^{3}}\, \,
\frac{\Gamma_{\rm R}}{M_{\rm R}}\,\, \frac{4m_\pi^2}{M_{\rm R}^2}  \, ,   \nonumber
\\
t^s_{2n}&=& \sum_R \frac{1}{\rho(M_{\rm R}^2)^{2n+3}}\, \,
\frac{\Gamma_{\rm R}}{M_{\rm R}}\,\, \left(\frac{m_\pi^2}{M_{\rm R}^2}\right)^n  \, ,
\qquad \mbox{for } n\geq 1\, .
\end{eqnarray}
The subscript $R$ denotes the resonances $R$ with the appropriate
$IJ$ quantum numbers of the channel. Only one multiplet of scalars
and vector mesons is considered in the present study.

Up to $\cO(p^4)$, the expansion of the $t$--channel resonance
exchange $\sum_R T^{\rm tR}(s)=\sum_n t^t_{2n} \left(\frac{s-4
m_\pi^2}{m_\pi^2}\right)^n$  yields
\begin{enumerate}
\item{$IJ=11$ channel}
\bqa
t_{0}^t&=& \left(   \frac{4\Gamma_{\rm S}}{9M_{\rm S}^3}+\frac{2\Gamma_{\rm V}}{M_{\rm V}^3}
\right)\, m_\pi^2
+   \left(    \frac{8\Gamma_{\rm S}}{3M_{\rm S}^5}+\frac{12\Gamma_{\rm V}}{M_{\rm V}^5}
\right)\, m_\pi^4,  \nonumber
\eqa

\bqa
t_2^t&=& \left(    \frac{\Gamma_{\rm S}}{9M_{\rm S}^3}+\frac{\Gamma_{\rm V}}{2M_{\rm V}^3}
\right)\, m_\pi^2
+   \left(   \frac{2\Gamma_{\rm S}}{9M_{\rm S}^5}+\frac{5\Gamma_{\rm V}}{M_{\rm V}^5}
\right)\, m_\pi^4,   \nonumber
\eqa

\bqa
t_4^t&=& \left(    \frac{-\Gamma_{\rm S}}{9M_{\rm S}^5}+\frac{\Gamma_{\rm V}}{2M_{\rm V}^5}
\right)\, m_\pi^4 .
\eqa

\item{$IJ=00$ channel}
\bqa
t_0^t&=&   \left(   \frac{-4\Gamma_{\rm S}}{3M_{\rm S}^3}+\frac{36\Gamma_{\rm V}}{M_{\rm V}^3}
\right) \, m_\pi^2
+   \left(\frac{-56\Gamma_{\rm S}}{9M_{\rm S}^5}+\frac{232\Gamma_{\rm V}}{M_{\rm V}^5}
\right)\, m_\pi^4 \, ,   \nonumber
\eqa
\bqa
t_2^t&=&   \left(   \frac{-\Gamma_{\rm S}}{3M_{\rm S}^3}+\frac{9\Gamma_{\rm V}}{M_{\rm V}^3}
\right)  \, m_\pi^2
+  \left(\frac{-2\Gamma_{\rm S}}{3M_{\rm S}^5}+\frac{42\Gamma_{\rm V}}{M_{\rm V}^5}
\right)\, m_\pi^4\, , \nonumber
\eqa
\bqa
t_4^t&=&  \left(  \frac{2\Gamma_{\rm S}}{9M_{\rm S}^5}-\frac{4\Gamma_{\rm V}}{M_{\rm V}^5}
\right)  \, m_\pi^4 .
\eqa

\item{$IJ=20$ channel}
\bqa
t_0^t&=&  -  \left(  \frac{4\Gamma_{\rm S}}{3M_{\rm S}^3}+\frac{18\Gamma_{\rm V}}{M_{\rm V}^3}
\right) \, m_\pi^2
-  \left( \frac{56\Gamma_{\rm S}}{9M_{\rm S}^5}+\frac{116\Gamma_{\rm V}}{M_{\rm V}^5}
\right) \, m_\pi^4 \, ,   \nonumber
\eqa

\bqa
t_2^t&=& -\left(  \frac{\Gamma_{\rm S}}{3M_{\rm S}^3}+\frac{9\Gamma_{\rm V}}{2M_{\rm V}^3}
\right) \, m_\pi^2
- \left(\frac{2\Gamma_{\rm S}}{3M_{\rm S}^5}+\frac{21\Gamma_{\rm V}}{M_{\rm V}^5}
\right)\, m_\pi^4 \, ,  \nonumber
\eqa

\bqa
t_4^t&=& \left(  \frac{2\Gamma_{\rm S}}{9M_{\rm S}^5}+\frac{2\Gamma_{\rm V}}{M_{\rm V}^5}
\right) \, m_\pi^4\, .
\eqa

\end{enumerate}

The quantities $t_0^t$ and $t^s_0$ in each channel actually  gives, respectively, the
crossed-channel and the $s$-channel resonance
contribution to the scattering length parameter.

\subsection{Chiral perturbation theory scattering amplitude}

In the large--$N_C$ limit, the $\chi$PT scattering amplitude is
given up to $\mathcal{O}(p^4)$ by the coefficients~\cite{chptms}:
\begin{enumerate}
\item{$IJ=11$ channel}

 \bqa
t_0^{\chi PT}&=&0, \nonumber \\
t_2^{\chi PT}&=& \frac{m_\pi^2}{96\pi f^2}-\frac{m_\pi^4}{6\pi f^4}\, L_3\, , \nonumber \\
t_4^{\chi PT}&=& \frac{-m_\pi^4}{24 \pi f^4}L_3\, , \nonumber \\
T(0)^{\chi PT}&=& \frac{-m_\pi^2}{24\pi f^2}\, .
\label{eq.chpt11}\eqa

\item{$IJ=00$ channel}
\bqa
t_0^{\chi PT}&=&\frac{7m_\pi^2}{32\pi f^2}+\frac{m_\pi^4}{2\pi f^4}(15L_2+5L_3-\frac{5}{2}L_5+5L_8)
\, , \nonumber \\
t_2^{\chi PT}&=& \frac{m_\pi^2}{16\pi f^2}+\frac{m_\pi^4}{\pi f^4}(5L_2+2L_3)\, , \nonumber \\
t_4^{\chi PT}&=& \frac{m_\pi^4}{24\pi f^4}(25L_2+11L_3)\, , \nonumber \\
T(0)^{\chi PT}&=& \frac{-m_\pi^2}{32\pi f^2}+\frac{m_\pi^4}{6\pi
f^4}(25L_2+11L_3-\frac{15}{2}L_5+15L_8) \, . \label{eq.chpt00} \eqa

\item for IJ=20 channel one has:
\bqa t_0^{\chi PT}&=&\frac{-m_\pi^2}{16\pi f^2}+\frac{m_\pi^4}{\pi
f^4}(3L_2+L_3-\frac{1}{2}L_5+L_8)\, ,
\nonumber \\
t_2^{\chi PT}&=& \frac{-m_\pi^2}{32\pi f^2}+\frac{m_\pi^4}{2\pi f^4}(4L_2+L_3) \, , \nonumber \\
t_4^{\chi PT}&=& \frac{m_\pi^4}{12\pi f^4}(5L_2+L_3)\, , \nonumber \\
T(0)^{\chi PT}&=& \frac{m_\pi^2}{16\pi f^2}+\frac{m_\pi^4}{3\pi
f^4}(5L_2+L_3-\frac{3}{2}L_5+3L_8) \,  . \label{eq.chpt20} \eqa
\end{enumerate}
where the $t_n^{\chi PT}$ are given by the threshold expansion of
the chiral amplitude $T(s)^{\chi PT} =\sum_{n=0}^\infty t_{2n}^{\chi
PT}\, \left(\frac{s-4 m_\pi^2}{m_\pi^2}\right)^n$ and  $T(0)^{\chi PT}$
denotes the value of the ${\chi PT}$ scattering amplitude at $s=0$.
The constant  $f$ is the chiral limit of the pion decay constant,
$f\approx 88$~MeV~\cite{chptoneloop}. In order to get the expressions in
Eqs.(\ref{eq.chpt11})--(\ref{eq.chpt20}), the one-loop contributions
have been dropped and we have made use of the large--$N_C$ relations
$L_4=L_6=0$ and $L_1= L_2/2$~\cite{chptms}.

\subsection{Matching dispersive and $\chi$PT
expressions}\label{match}

Having obtained the resonance expansions as well  as the chiral
expansions  at threshold, matching conditions can be set up between
the two kind of amplitudes. For simplicity we in the following only
introduce minimal set of resonances, i.e., only $\sigma$ and $\rho$.
We point out that in case of need it is
 straightforward to add higher resonances in the present scheme.

The matching in Eq.~(\ref{eq.Tmatch}),  considered order by order  in the threshold
expansion,  leads to a series of relations. Only the terms up to
$\mathcal{O}(p^4)$ in the chiral expansion are retained in this work:
\begin{enumerate}

\item{$IJ=11$ channel}
\bqa \label{sum110}
&& \frac{1}{24\pi f^2}=
\frac{4\Gamma_{\rm S}}{9M_{\rm S}^3}+\frac{6\Gamma_{\rm V}}{M_{\rm V}^3}
+  \left(  \frac{8\Gamma_{\rm S}}{3M_{\rm S}^5}
+\frac{36\Gamma_{\rm V}}{M_{\rm V}^5}   \right) \, m_\pi^2   \, ,
\eqa

\bqa  \label{sum111}
&& \frac{1}{96\pi f^2}-\frac{m_\pi^2}{6\pi f^4}L_3=
\frac{\Gamma_{\rm S}}{9M_{\rm S}^3}+\frac{3\Gamma_{\rm V}}{2M_{\rm V}^3}
+  \left(\frac{2\Gamma_{\rm S}}{9M_{\rm S}^5}
+\frac{15\Gamma_{\rm V}}{M_{\rm V}^5}    \right) \, m_\pi^2   ,  \nonumber \\
\eqa

\bqa\label{sum112} &&-\, \frac{L_3}{24\pi f^4}\, =\, -\,
\frac{\Gamma_{\rm S}}{9M_{\rm S}^5}\, +\, \frac{3\Gamma_{\rm
V}}{2M_{\rm V}^5} , \eqa

\item{$IJ=00$ channel}

\bqa  \label{sum000} &&\hspace*{-1.5cm}  \frac{1}{4\pi f^2}+\frac{m_\pi^2}{3\pi
f^4}(10L_2+2L_3)=
\frac{8\Gamma_{\rm S}}{3M_{\rm S}^3}+\frac{36\Gamma_{\rm V}
}{M_{\rm V}^3}
+  \left(   \frac{160\Gamma_{\rm S}}{9M_{\rm S}^5}
+\frac{232\Gamma_{\rm V}}{M_{\rm V}^5}  \right)\, m_\pi^2 \, ,  \nonumber \\
\eqa

\bqa \label{sum001} &&\frac{1}{16\pi f^2}+\frac{m_\pi^2}{\pi
f^4}(5L_2+2L_3)=
\frac{2\Gamma_{\rm S}}{3M_{\rm S}^3}+\frac{9\Gamma_{\rm V}}{M_{\rm V}^3}
+   \left(  \frac{28\Gamma_{\rm S}}{3M_{\rm S}^5}
+\frac{42\Gamma_{\rm V}}{M_{\rm V}^5}\right) \, m_\pi^2  \, , \nonumber \\
\eqa

\bqa \label{sum002} \frac{1}{24\pi f^4}(25L_2+11L_3)\, =\,
\frac{11\Gamma_{\rm S}}{9M_{\rm S}^5}\, -\, \frac{4\Gamma_{\rm
V}}{M_{\rm V}^5}, \eqa

\item{$IJ=20$ channel}

 \bqa \label{sum200}
 && \hspace*{-2cm}  -\, \frac{1}{8\pi f^2}\, +\, \frac{m_\pi^2}{3\pi
f^4}(4L_2+2L_3)=
  -\, \frac{4\Gamma_{\rm S}}{3M_{\rm S}^3} \, -\, \frac{18\Gamma_{\rm V}}{M_{\rm V}^3}\,
  -\, \left(   \frac{56\Gamma_{\rm S}}{9M_{\rm S}^5}
+\frac{116\Gamma_{\rm V}}{M_{\rm V}^5}   \right) \, m_\pi^2 \, , \nonumber \\
\eqa

\bqa \label{sum201} &&
 \hspace*{-2cm}   -\, \frac{1}{32\pi f^2}
 \, +\, \frac{m_\pi^2}{2\pi f^4}(4L_2+L_3)= 
    -\, \frac{\Gamma_{\rm S}}{3M_{\rm S}^3}   \, -\,  \frac{9\Gamma_{\rm V}}{2M_{\rm V}^3}   \,
    -\,  \left(   \frac{2\Gamma_{\rm S}}{3M_{\rm S}^5}
+\frac{21\Gamma_{\rm V}}{M_{\rm V}^5}    \right) \, m_\pi^2  \,  , \nonumber \\
\eqa

\bqa \label{sum202}  \frac{(5L_2+L_3)}{12\pi f^4}=\,
\frac{2\Gamma_{\rm S}}{9M_{\rm S}^5}\, +\,  \frac{2\Gamma_{\rm
V}}{M_{\rm V}^5} \, . \eqa

\end{enumerate}
A global factor $m_\pi^2$  has been simplified in
Eqs.~(\ref{sum110}), (\ref{sum111}), (\ref{sum000}), (\ref{sum001}),
(\ref{sum200}) and (\ref{sum201}), and Eqs.~(\ref{sum112}),
(\ref{sum002}) and (\ref{sum202}) have been divided by a factor
$m_\pi^4$.
 Notice that the matching equations do not depend
explicitly on the low-energy couplings $L_5$ and $L_8$. The
contribution from the $L_5$ $\pi^4$ operator to the scattering
amplitude is canceled out up to a constant term by the $L_5$ part of
the pion wave function renormalization $Z_\pi$  of the external
legs. The $L_8$ operator does not contain derivatives and it just
adds another energy independent term to the $\pi\pi$--amplitude.
Since the constant contributions vanish when considering the
difference $T(4m_\pi^2)-T(0)$ (with $T(4m_\pi^2)=t_0$ in our
notation), $L_5$ and $L_8$ do no longer appear explicitly in the
matching equations.

The first thing to notice is that the identities related to the
matching  $t_0^{\chi PT}-T^{\chi PT}(0) = t_0^s + t_0^t$
(Eqs.(\ref{sum110}),~(\ref{sum000}) and (\ref{sum200})) are linear
combinations of the other two matching relations for $t_2^{\chi PT}$
and $t_4^{\chi PT}$.  This is due to the fact that $T^{\chi
PT}(s)-T^{\chi PT}(0)$ vanishes at zero by construction. Hence, its
threshold expansion carries the implicit relation $t_0^{\chi
PT}-T^{\chi PT}(0)=4 t_2^{\chi PT}- 16 t_4^{\chi PT}$ in our notation.

The physical widths and masses, $\Gamma_{\rm R}$ and $M_{\rm R}$,  carry an
implicit dependence on $m_\pi^2$, which can be expressed in the form
\begin{equation}
\frac{\Gamma_{\rm R}}{M_{\rm R}^3}\, =\, \frac{\Gamma_{\rm R}^{(0)}}{M_{\rm R}^{(0)\,
3}}\,\left[\, 1 \, +\, \alpha_R\, \frac{m_\pi^2}{M_{\rm R}^{(0)\,
2}}\,\,\, +\, \, \, \mathcal{O}\left(m_\pi^4\right)\, \right]\, .
\end{equation}
The constants $M_{\rm R}^{(0)}$ and $\Gamma_{\rm R}^{(0)}$ are respectively the
mass and width of the resonance $R$ in the chiral limit and
$\alpha_R$ parameterizes the deviation from the chiral limit.

The matching to $\chi$PT at $\cO(p^2)$ is given by the
$\cO(m_\pi^0)$ terms in Eqs.~(\ref{sum111}), (\ref{sum001}) and
(\ref{sum201}).  The three different channels produce the same
equation,
\begin{equation}
\label{sumksrf} \frac{1}{16 \pi
f^2}=\frac{9\Gamma^{(0)}_V}{M_{\rm V}^{(0)\,
3}}+\frac{2\Gamma^{(0)}_S}{3M_{\rm S}^{(0)\, 3}},
\end{equation}
which is nothing but a extension to the well known KSRF
relation~\cite{ksrf00}.

One old way to express the KSRF relation is
the following,
 \be\label{KSRF1}
  g^2_{\rho\pi\pi}=\frac{M_\rho^2}{2f_\pi^2}\ ,
 \ee
where $g_{\rho\pi\pi}$ characterizes the $\rho-\pi\pi$ coupling.
For a massive Yang-Mills model, the chiral limit of the $\rho$ width is given by
 \be\label{KSRF2}
\Gamma_\rho=\frac{g^2_{\rho\pi\pi}}{48\pi} M_\rho\ .
 \ee
Combining Eqs.~(\ref{KSRF1}) and (\ref{KSRF2}) leads to
 \be\label{KSRF3}
  \frac{1}{16 \pi
f^2}=\frac{6\Gamma^{(0)}_V}{M_{\rm V}^{(0)\, 3}}\ .
 \ee
The difference  between Eqs.~(\ref{sumksrf}) and (\ref{KSRF3}) on
the $r.h.s.$ is clearly understood when we examine the matching in
the IJ=11 channel: it comes from the crossed channel vector and
scalar meson exchanges, which is absent in Eq.~(\ref{KSRF3}).
Furthermore, it is remarkable to notice that, all the three channels
lead to the same generalized KSRF relation. The modification of the
KSRF relation due to the crossed channel resonance exchange was
first noticed in Ref.~\cite{ksrf0}~\footnote{Instead of
Eq.~(\ref{KSRF1}), the relation given in Ref.~\cite{ksrf0} is,
$g^2_{\rho\pi\pi}=\frac{M_\rho^2}{3f_\pi^2}$. In Ref.~\cite{igi},
Hikasa and Igi included scalar exchange  and were able to obtain a
relation similar to Eq.~(\ref{sumksrf}) in all three channels,
assisted by N/D method. }. Our work stressed that the correct
expression of the so-called KSRF relation can be obtained in a
systematic way without relying on any particular lagrangian
formalism: once subtracted partial wave dispersion relations
combined with chiral symmetry and large $N_C$ expansion (or narrow
width approximation) generates our modified KSRF relation. The
matching at both high and low energies are crucial for establishing
this constraint. The different contributions to the KSRF relation
are summarized in Table~\ref{tab.KSRF}.
\begin{table}[!t]
\begin{center}
\begin{tabular}{|c|c|c|c|c|}
  \hline
\rule[-0.7em]{0em}{1.9em}
   & $T(0)$ & $t_0^{\rm tR}$ & $t_0^{\rm sR}$ & $t_0^{\chi PT}$ \\
  \hline
\rule[-0.7em]{0em}{1.9em}
  $IJ=11$ & $-\frac{m_\pi^2}{24\pi f^2}$ & $\frac{4\Gamma_S}{9M_S^3}+\frac{2\Gamma_V}{M_V^3}$
        & $\frac{4\Gamma_V}{M_V^3}$ & 0 \\
  \hline
\rule[-0.7em]{0em}{1.9em}
  $IJ=00$ & $-\frac{m_\pi^2}{32\pi f^2}$ & $-\frac{4\Gamma_S}{3M_S^3}+\frac{36\Gamma_V}{M_V^3}$
        & $\frac{4\Gamma_S}{M_S^3}$ & $\frac{7m_\pi^2}{32\pi f^2}$\\
  \hline
\rule[-0.7em]{0em}{1.9em}
  $IJ=20$ & $\frac{m_\pi^2}{16\pi f^2}$ & $-\frac{4\Gamma_S}{3M_S^3}-\frac{18\Gamma_V}{M_V^3}$
        & 0 & $-\frac{m_\pi^2}{16\pi f^2}$ \\
  \hline
\end{tabular}
\caption{{\small
Summary of the different contributions  $T(0)$, $t_0^{\rm tR}$, $t_0^{\rm SR}$
to the scattering lengths at leading order in the
$m_\pi^2$ expansion.  The generalized KSRF-relation derives from the matching of the
sum of the first three columns to the
$\chi$PT prediction, $t_0^{\chi PT}$.
In the last line,  $T(0)$ contains the sum of $-|T(0)|$ and the $IJ=20$
virtual pole contribution.
}}  \label{tab.KSRF}
\end{center}
\end{table}

The \  matching \ to \ $\chi$PT \ at \ $\cO(p^4)$, \ gives \ another
\  six  \ identities. \ The \  ${\cO\left((s-4m_\pi^2)^2\right)}$
terms from the $IJ=11,00,20$ channels (Eqs.(\ref{sum112}),
(\ref{sum002}) and (\ref{sum202})) provide the constraints
\begin{eqnarray}
\label{sumL2} L_2\, &=& \, 12\pi f^4\frac{\Gamma_{\rm V}^{(0)}}{M_{\rm V}^{(0)\,
5}} \, ,
\\
\label{sumL3} L_3 &=& 4 \pi f^4
\left(\frac{2\Gamma_{\rm S}^{(0)}}{3M_{\rm S}^{(0)\,
5}}-\frac{9\Gamma_{\rm V}^{(0)}}{M_{\rm V}^{(0)\, 5}}  \right)\, .
\end{eqnarray}
The Eqs.~(\ref{sumL2}), (\ref{sumL3})  provide a large $N_C$
prediction for the LECs $L_2$ and $L_3$. The two expressions
obey the positivity
constraints: $L_2>0$ and $3L_2+L_3>0$ as revealed in
Ref.~\cite{pham}.

 The remaining $\cO(p^4)$  relations are provided by the
$\cO(m_\pi^2)$ terms in the ${\cO(s-4 m_\pi^2)}$ equations
(Eqs.(\ref{sum111}), (\ref{sum001}) and (\ref{sum201})), and produce
\begin{eqnarray}
\label{sumalpha} 0&=& \frac{2}{3}\,\,
\frac{\Gamma_{\rm S}^{(0)}}{M_{\rm S}^{(0)\, 5}}\, \left[\alpha_S+6\right] \,
+\, \frac{ 9\, \Gamma_{\rm V}^{(0)}}{M_{\rm V}^{(0)\, 5}} \,
\left[\alpha_V+6\right]\, .
\end{eqnarray}
The novel relation, Eq.~(\ref{sumalpha}) casts an interesting
relation between resonance parameters. The Eqs.~(\ref{sumL2}), (\ref{sumL3}),~(\ref{sumalpha})
and the extended KSRF relation,
Eq.~(\ref{sumksrf}) are generated simultaneously, in a systematic
way, by a matching to $\chi$PT amplitude at different chiral orders.
The following section is devoted to a better understanding to the
new relation, Eq.~(\ref{sumalpha}).

\section{On the consistency of lagrangian
models}\label{examine}

In this section, we inspect several
phenomenological lagrangians that have been proposed in order to
describe the resonance interactions. Firstly, we will consider the
toy model with a linear sigma meson representation
and the chiral gauged model~\cite{donoghue}, which only introduces vector
mesons. These examples illustrate very nicely the
expected properties that a meson theory must fulfill.  A similar analysis
can be also carried within the hidden local symmetry model~\cite{HLS}.
We end the section with an extensive analysis within resonance chiral
theory~\cite{Ecker89,rcht-op6}.

\subsection{Linear sigma  model}

The linear sigma model (L$\sigma$M) with massive pions is given by
the lagrangian
\begin{equation}
\mathcal{L}_{L\sigma M}\, =\, \frac{1}{2}\left[(\partial \pi)^2 +
(\partial \sigma)^2\right]   \, + \, \frac{1}{2}
\mu^2\left[\pi^2+\sigma^2\right] \, - \, \frac{1}{4}\lambda\,
\left[\pi^2 + \sigma^2\right]^2  \, +\,  f\,  m_\pi^2 \, \sigma\, ,
\end{equation}
with $f=\sqrt{\frac{\mu^2}{\lambda}}$. No vectors are considered in
this model.

After \ shifting \ the  \ $\sigma$  \ field  \ due \  to \  its  \ vacuum  \ expectation  \ value \
${\langle \sigma \rangle
=\sqrt{\frac{\mu^2}{\lambda}}\, \left(1\, + \,\frac{m_\pi^2}{\mu^2}+...\right)}$,  one
gets the tree-level mass term
\begin{equation}
M_\sigma^2\, = \, M_\sigma^{(0)\, 2}\, \left[\, 1 \, + \, \frac{3\,
m_\pi^2}{M_\sigma^{(0)\, 2}}\, + \, ...\right]\, ,
\end{equation}
with $M_{\sigma}^{(0)\, 2}=2\, \mu^2$. The large-$N_C$ width is given by
the $\sigma-\pi\pi$ vertex:
\begin{equation}
\Gamma_\sigma\, = \, \Gamma_\sigma^{(0)}\, \left[\, 1 \, -  \,
\frac{3 \, m_\pi^2}{2\, M_\sigma^{(0)\, 2}}\, + \, ...\right]\, ,
\end{equation}
with $\Gamma_\sigma^{(0)}=  \frac{3\lambda}{16\pi} M_\sigma^{(0)} $.
Putting both expressions together in the combination
$\Gamma_\sigma/M_\sigma^3$ one gets
\begin{equation} \alpha_S\, =\,
\left[\Frac{M_\sigma^5}{\Gamma_\sigma}\Frac{d}{dm_\pi^2}
\left(\Frac{\Gamma_\sigma}{M_\sigma^3}\right)\right]_{m_\pi^2=0}
   \, =\, -6 \, .
\end{equation}
Since there are no vectors in the theory, Eq.~(\ref{sumalpha}) is
exactly fulfilled.  Likewise, the L$\sigma$M produce the value
\begin{equation}
\Frac{ \Gamma_\sigma^{(0)}  }{M_\sigma^{(0)\,\, 3} }
\, =\,  \Frac{3 \, \lambda}{ 32 \pi \mu^2} \,= \,  \frac{3}{32\pi f^2}\, .
\end{equation}
Since there are no vectors in the theory, this result fulfills the modified--KSRF
relation in Eq.~(\ref{sumksrf}) for any value of the couplings
$\mu$ and $\lambda$.

This can be better understood through the explicit diagramatic
calculation. The analysis of the $\pi\pi$--scattering amplitude
shows that the structure of the L$\sigma$M lagrangian  ensures a
good high energy behaviour, independently of the value of the
resonance parameters. Since the model obeys the proper high and low
energy limits by construction, no resonance constraint can be
extracted, just the usual low-energy coupling determinations for
$L_2$ and $L_3$.

This exercise shows how, in order to fulfill the former constraints,
a theory must have a right asymptotic behavior at high and low
energies. In this case, chiral invariance ensures the right low
energy properties and the L$\sigma$M renormalizability ensures the
proper high energy asymptotic behavior.
However, the next example shows that renormalizability is not actually the
necessary condition for the fulfillment of  our large--$N_C$ sum-rules.

\subsection{The gauged chiral model}
In this model, vector and axial-vector resonances are included as
gauge bosons in the $SU(2)$ $\chi$PT lagrangian~\cite{donoghue}:
\begin{eqnarray}
\mathcal{L}_{G\chi M}&= & \frac{f_0^2}{4}  \langle D_\mu U D^\mu U^+
\rangle  \, +\, \frac{m_\pi^2 \, f^2}{4}   \langle U+U^\dagger
\rangle \, - \frac{1}{4} \langle
L_{\mu\nu}L^{\mu\nu}+R_{\mu\nu}R^{\mu\nu} \rangle \nonumber
\\ && \qquad +
M_0^2  \langle L_\mu L^\mu+R_\mu R^\mu  \rangle + B \langle L_\mu U
R^\mu U^+  \rangle  ,
\end{eqnarray}
with $\langle...\rangle$ short for trace in flavor space. The chiral
tensors are defined as
\begin{eqnarray}
U&=&\exp{\left(i\frac{\tau^a\pi^a}{f} \right)} ,\nonumber \\
D_\mu U&=&\partial_\mu U -i g L_\mu U + i g U R_\mu , \nonumber \\
L_\mu &=& \frac{\tau^a}{2}(V_\mu^a + A_\mu^a) , \nonumber \\
R_\mu &=& \frac{\tau^a}{2}(V_\mu^a - A_\mu^a) , \nonumber \\
L_{\mu\nu}&=& \partial_\mu L_\nu - \partial_\nu L_\mu -i g [L_\mu , L_\nu], \nonumber \\
R_{\mu\nu}&=& \partial_\mu R_\nu - \partial_\nu R_\mu -i g [R_\mu , R_\nu], \nonumber \\
\end{eqnarray}
where $V_\mu^a$ and $A_\mu^a$ are the $SU(2)$ vector and
axial-vector triplets, respectively, $\rho$ and $a_1$, and $\tau^a $
are the Pauli matrices. The last term, with coefficient $B$, is not essential and
allows the model to be reasonably compatible with phenomenology~\cite{donoghue}.
It is dropped off in our analysis, following the derivation in the original paper.

The calculation the
tree level $\rho\rightarrow\pi\pi$ decay width  and
the low-energy $\pi\pi$--scattering amplitude casts,
\begin{eqnarray}
\Gamma_{\rho\rightarrow\pi\pi} &=&
\frac{g_\rho^2\, M_\rho}{48\pi } \, \rho(M_\rho^2)^3\,  ,
\label{eq.GammaGChM}
\\
L_2 &=& \Frac{g_\rho^2  f^4}{4 \, M_\rho^4}\, ,
\label{eq.L2GChM}
\\
L_3 &=& \Frac{3\, g_\rho^2 f^4}{4\, M_\rho^4}\, ,
\label{eq.L3GChM}
\end{eqnarray}
where the parameters  $g_\rho$, $f$, $M_\rho$
are related to the original couplings in the lagrangian
through\footnote{Notice
the missprint in the original paper~\cite{donoghue}, where
the authors refer $M_\rho$ instead of $M_0$
in the relations for $g_\rho$ and $f$  at Eqs.~(\ref{eq.grho})--(\ref{eq.fpi}).}
\begin{eqnarray}
g_\rho^2  &=&   g^2 \, \left(1\, - \, \Frac{g^2 f^2}{4 M_0^2}\right)^2\, ,
\label{eq.grho}
 \\
f^2 &=& f_0^2 \, \left(1 \, +\,  \Frac{g^2 f_0^2}{2 \, M_0^2}\right)^{-1}\, ,
 \label{eq.fpi}
\\
M_\rho^2 &=& 2 M_0^2\, .
\end{eqnarray}
The difference between the pion decay constant $f$ and the coupling
$f_0$ is due to the presence of $\pi-A_1$ mixing terms in the gauge
chiral model lagrangian. A similar thing happens with the coupling
$g$ and the effective $\rho-\pi\pi$ parameter $g_\rho$. By means of
Eq.~(\ref{eq.GammaGChM}) one gets $\Gamma_\rho^{(0)}= g_\rho^2
M_\rho /48\pi$ and then it is not difficult to realize that the
corresponding low-energy couplings in
Eqs.~(\ref{eq.L2GChM})--(\ref{eq.L3GChM}) exactly agree our sum-rule
predictions in Eqs.~(\ref{sumL2})--(\ref{sumL3}).

The parameters $M_\rho$, $f$, $g_\rho$ are independent of the pion mass
at large--$N_C$ and,  hence, the $\alpha_V$ corresponding to the
gauge chiral model is given by
\begin{equation}
  \alpha_V  \, = \, \left[ \Frac{M_\rho^5}{\Gamma_\rho}\,
  \Frac{d}{dm_\pi^2}\left(\Frac{\Gamma_\rho}{M_\rho^3}\right)\right]_{m_\pi^2=0}
  \, =\, -6 \, .
\end{equation}
Since there are no
scalars in the theory, the relation in Eq.~(\ref{sumalpha}) is
trivially obeyed for any value of $M_\rho$, $g_\rho$ and $f$,  and no
resonance constraint is extracted.

This illustrates that
renormalizability is not a necessary condition for the fulfillment
of our resonance constraints. The key-point is that
the amplitudes must obey a proper high energy behavior. The inspection of the
$IJ=11$ $\pi\pi$--scattering amplitude at  $s\to \infty$
yields,
\begin{equation}
T^1_1(s)\, =\, \Frac{s}{96\pi f^2}\,
\left[1\, -\, \Frac{3 \, g_\rho^2 f^2}{M_\rho^2}\right] \, \, +\, \, \,
\cO(s^0)\, .
\end{equation}
Although one could  a priori  expect the presence of $\cO(s\,
m_\pi^2)$ terms, they disappear from the amplitude after precise
cancelations between different contributions. The absence of these
terms explains why our $\alpha_V$ relation in Eq.~(\ref{sumalpha})
is trivially obeyed and produces no constraint on the resonance
couplings. Moreover, by demanding that the $\cO(s)$ term vanishes
one gets $3 g_\rho^2 f^2/M_\rho^2=1$, which is nothing else but the
KSRF relation in Eq.~(\ref{sumksrf}) in the absence of scalars. The
analysis of the $IJ=00$ and $IJ=20$ channels gives identical
results.

\subsection{Minimal Resonance Chiral Theory}

In the original \  Resonance Chiral Theory lagrangian (R$\chi$T)
proposed \ in Ref.~\cite{Ecker89}, the authors built the most general
chiral invariant lagrangian that contributed at low energies to the
$\cO(p^4)$ $\chi$PT couplings. For sake of this, just operators with
at most one resonance field were considered:
\begin{eqnarray}
\mL_V &=& \frac{F_V}{2\sqrt{2}}\langle V_{\mu\nu}f_+^{\mu\nu}\rangle
\, + \, \frac{ i G_V}{2\sqrt{2}}\langle V_{\mu\nu}
[u^\mu,u^\nu]\rangle \, ,
\\
\mL_S &=& c_d \langle S u^\mu u_\mu\rangle \, +\, c_m\langle S
\chi_+\rangle\, ,
\end{eqnarray}
and the kinetic terms
\begin{eqnarray}
\mL_V^{\rm Kin} &=& - \frac{1}{2}\langle \nabla^\mu V_{\mu\lambda}
\nabla_\nu V^{\nu\lambda}\rangle \, + \, \frac{1}{4} M_{\rm V}^2 \langle
V_{\mu\nu} V^{\mu\nu}\rangle \, ,
\\
\mL_S^{\rm Kin} &=&  \frac{1}{2}\langle \nabla^\mu S\nabla_\mu
S\rangle \, -\, \frac{1}{2}M_{\rm S}^2 \langle SS\rangle \, ,
\end{eqnarray}
where the chiral tensors $u^\mu \sim p_\pi$, $\chi_+ \sim m_\pi^2$
and $f_{+}^{\mu\nu}$ containing external vector and axial-vector
sources are defined in Ref.~\cite{Ecker89}. The spin-1 fields are
given in the antisymmetric tensor formalism. The resonance masses did
not depend on the quark masses in the original approach.

The vector width was found to be
\begin{equation}
\Gamma_{\rm V} \, =\, \Gamma_{\rm V}^{(0)}\, \left[\, 1\, + \,
\frac{m_\pi^2}{M_{\rm V}^2}\left(-6 - \frac{16 c_d c_m M_{\rm V}^2}{f^2 M_{\rm S}^2}
\right) \, + \,...\right]\, ,
\end{equation}
with the chiral limit of the  $\rho\to \pi\pi$ width,
\begin{equation}
\Gamma_{\rm V} ^{(0)}=\frac{G_V^2 M_{\rm V}^3}{48 \pi f^4}\, .
\end{equation}
The first term in the $m_\pi^2$ correction comes from the $V\pi\pi$
vertex and the width phase-space factor $\rho(M_V^2)$. The second term, proportional to
$c_d c_m/M_{\rm S}^2$, comes from the pion wave function renormalization
at large--$N_C$. It appears for $m_q\neq 0$ due to the coupling of
the isosinglet resonances to the vacuum through the operator
$c_m\langle S \chi_+\rangle$~\cite{Kpi-SFF,FpiJJSC}.

The corresponding  scalar width is
\begin{equation}
\Gamma_{\rm S} \, =\, \Gamma_{\rm S} ^{(0)}\, \left[\, 1\, + \,
\frac{m_\pi^2}{M_{\rm S}^2}\left(-6 +\frac{4 c_m}{c_d} - \frac{16 c_d c_m
}{f^2}\right) \, + \,...\right]\, ,
\end{equation}
with the $\sigma \to \pi\pi$ width in the chiral limit,
\begin{equation}
\Gamma_{\rm S}^{(0)}=\frac{3 c_d^2 M_{\rm S}^3}{16\pi f^4}\, .
\end{equation}
When we refer to $\sigma$, we denote the $SU(2)$ singlet
$\sigma=\sqrt{\frac{2}{3}}S_0-\sqrt{\frac{1}{3}} S_8  \sim
\frac{1}{\sqrt{2}}(\bar{u}u +\bar{d}d)$. The first term in the
$m_\pi^2$ correction is produced by the $S\pi\pi$ vertex in the
$c_d\langle S u_\mu u^\mu\rangle$ operator and the width phase-space factor $\rho(M_S^2)$.
The second contribution
is produced by the $S\pi\pi$ vertex from the $c_m\langle
S\chi_+\rangle $ operator. Finally, the last term, proportional to
$c_d c_m/M_{\rm S}^2$, comes from the pion wave function renormalization
and it is also utterly linked to the $c_m\langle S \chi_+\rangle$
operator.

Substituting the widths  \ provided by the minimal
R$\chi$T~\cite{Ecker89} \ into the modified-KSRF relation of
Eq.(\ref{sumksrf}) one gets
\begin{equation}
1\, \, =\, \, \Frac{2 c_d^2}{f^2}\, \, + \, \, \Frac{3 G_V^2}{f^2}\,
.
\label{eq.sumKSRF-RChT}
\end{equation}

Since the R$\chi$T is explicitly chiral invariant, at low energies
one recovers the $\chi$PT structure independently of the value of the resonance
parameters $M_{\rm V}$, $M_{\rm S}$, $G_V$, $c_d$, $c_m$.  Eqs.(\ref{sumL2})
and (\ref{sumL3}) leads to the low energy coupling determinations
\begin{eqnarray}
L_2 &=&  12\pi f^4\frac{\Gamma_{\rm V}^{(0)}}{M_{\rm V}^{(0)\, 5}}
\, =\, \frac{ G_V^2}{ 4 M_{\rm V}^2} \, ,\label{L2sr}
\\
L_3  &=& 4 \pi f^4   \left(\frac{2\Gamma_{\rm S}^{(0)}}{3M_{\rm
S}^{(0)\, 5}}-\frac{9\Gamma_{\rm V}^{(0)}}{M_{\rm V}^{(0)\, 5}}
\right) \, =\, - \frac{3 G_V^2}{ 4 M_{\rm V}^2}  + \frac{c_d^2}{2
M_{\rm S}^2}\, ,\label{L3sr}
\end{eqnarray}
in complete agreement with the expressions from the explicit
integration of the heavy resonances in the R$\chi$T
action~\cite{Ecker89}.

The chiral corrections to the ratios $\Gamma_{\rm R}/M_{\rm R}^3$  take the form
\begin{equation}
\alpha_V\, = \, -6 - \frac{16 c_d c_m M_{\rm V}^2}{f^2 M_{\rm S}^2} \, , \qquad
\qquad \alpha_S\, =\, -6 +\frac{4 c_m}{c_d} - \frac{16 c_d c_m
}{f^2}\, .
\end{equation}
The substitution of these values in Eq.(\ref{sumalpha}) leads to the
constraint
\begin{equation}
 \left[1\, -\, \Frac{4 c_d^2}{f^2}\right]\, c_m  \, =\, \Frac{6 \, G_V^2}{f^2}\, c_m\, ,
\end{equation}
leading \  to the upper bound  \  $G_V^2\leq f^2/6$. \  This is in \
contradiction \  with  \ the \  phenomenological value of the vector
coupling, which is found to be ${G_V^2/f^2\sim 0.5
}$~\cite{spin1fields}.

It is remarkable that all the problem is originated by
one single operator, $c_m\langle S\chi_+\rangle $. In the absence of
this term ($c_m=0$), one has  $\alpha_S=\alpha_V=-6$ and
Eq.(\ref{sumalpha}) is trivially fulfilled. This means that we
cannot  just add this single operator to the lagrangian. It must be
accompanied by extra appropriate terms.

\subsection{Extended Resonance Chiral Theory}

The study of  three-point QCD Green-functions at short distances has
shown that the original lagrangian is insufficient~\cite{3point}.
Problems have also arisen in the analysis at next-to-leading order
in $1/N_C$~\cite{nloNC}.  In general, a lagrangian made of operators
including just one resonance field produces wrong growing behaviors
of the amplitudes at high energies, inconsistent with perturbative
QCD and the operator product expansion~\cite{OPE}. During   recent
years,  different groups have worked on the development of
lagrangian including operators with two and three resonance
fields~\cite{3point,MVsplit,prepara}. A final compilation
of this operators can be found in Ref.~\cite{rcht-op6}.

We firstly focus ourselves on the
scalar sector of the theory. The relevant operators for the scalar mass and width
are~\cite{rcht-op6,prepara}
\begin{eqnarray}
\mL_{S}&=& \lambda_6^{S}\langle S \{\chi_+ ,u^\mu u_\mu\}\rangle \,
+\,\lambda_7^{S}\langle S u^\mu \chi_+ u_\mu\rangle \, ,
\\  \nonumber\\
\mL_{SS}&=& \lambda_1^{SS}\langle SS u^\mu u_\mu\rangle \, +\,
\lambda_2^{SS}\langle S u^\mu S u_\mu\rangle \, +\,
\lambda_3^{SS}\langle SS\chi_+\rangle \, ,
\\ \nonumber \\
\mL_{SSS} &=& \lambda_0^{SSS} \langle SSS\rangle \, +\,
\lambda_1^{SSS}\langle S\nabla^\mu S\nabla_\mu S\rangle \, .
\end{eqnarray}

In the scalar sector, the presence of the operator $c_m\langle S
\chi_+\rangle$ in the lagrangian induces non-zero vacuum expectation
value of the isosinglet field proportional to the quark masses. For
non-zero quark masses, one needs to perform the shift
$S=\overline{S}+ 4 c_m B_0 \mM/M_{\rm S}^{(0)\, 2}$, with  $M_{\rm S}^{(0)}$ the
scalar mass in the chiral limit and $\mM$ the quark mass matrix.
An alternative covariant shift
would be $S=\overline{S}+c_m \chi_+/M_{\rm S}^{(0)\, 2}$ but the former
one is more convenient for our calculation. This induces  a
wave-function renormalization of the pion and scalar fields,
${\pi=Z_\pi^\frac{1}{2} \pi^r}$ and $\overline{S}=Z_S^\frac{1}{2}
S^r$, respectively.

For the  large--$N_C$ analysis of the $\pi\pi$--scattering, we can
restrict ourselves to the $U(2)$ sector of the theory and work
within the isospin limit.  Hence, the relevant operators for the
mass and width $\Gamma[S^r\to \pi^r\pi^r]$ of the $U(2)$--isosinglet
scalar are given up to order $m_\pi^2$ by
\begin{equation}
\Delta \mL  =   -\, \frac{1}{2}M_{\rm S}^{{\rm eff}\,\,2}\, \langle S^r
S^r \rangle \, +\, c_d^{\rm eff}\, \langle S^r u^\mu u_\mu \rangle\,
+ \, c_m^{\rm eff} \, \langle S^r \chi_+\rangle  \, ,
\end{equation}
with the $m_\pi^2$ dependent parameter,
\begin{eqnarray}\label{cdeff}
%
%
c_d^{\rm eff} &=& c_d\, \left[ 1+ \delta c_d\, \Frac{m_\pi^2}{M_{\rm S}^{(0)\,
2}}\right]\, ,
\end{eqnarray}
given by the correction
\begin{equation}
\delta c_d\, = \,   \Frac{ 2 M_{\rm S}^{(0)\, 2}}{c_d}  (2 \lambda_6^S+\lambda_7^S)
+ \Frac{ 4 c_m}{c_d}
(\lambda_1^{SS}+\lambda_2^{SS})-2\lambda_1^{SSS}c_m \, .
\end{equation}
The $\cO(m_\pi^2)$ terms in $c_m^{\rm eff}=c_m\, \left[1+\cO(m_\pi^2)\right]$ and
$M_{\rm S}^{\rm eff}=M_{\rm S}^{(0)}\, \left[1+\cO(m_\pi^2)\right]$ are not relevant for
our problem since they contribute to the ratio $\Gamma_{\rm S}/M_{\rm S}^3$ at order $m_\pi^4$.
The pion decay constant up to $\cO(m_\pi^2)$ is provided in the
large--$N_C$ limit by~\cite{Ecker89,FpiJJSC}
\begin{equation}
%
%
%
%
f_\pi \, \, =\, \,f\,  Z_\pi^{-\frac{1}{2}} \, =\, f\, \left[1\, +\, \delta f\,
\Frac{m_\pi^2}{M_{\rm S}^{(0)\, 2}}\right]
\, , \qquad \qquad \mbox{with} \quad
\delta f\, = \, \Frac{4 c_d c_m}{f^2}\, .
\end{equation}
In what follows, we will denote the mass $M_{\rm S}^{\rm eff}$ simply as
$M_{\rm S}$, keeping  $M_{\rm S}^{(0)}$ for its chiral limit.

The relevant quantities in our KSRF
relations in Eqs.(\ref{sumksrf}) and (\ref{sumalpha}) are the ratios
$\Gamma/M^3$. In the scalar case, one finds
\begin{eqnarray}
\Frac{\Gamma_{\rm S}}{M_{\rm S}^3}\, &=& \, \Frac{3 \, c_d^{{\rm
eff}\, 2}  \, \rho(M_{\rm S}^2)}{16\pi \, f_\pi^4}\, \left[1 \, + \,
\Frac{4 m_\pi^2}{M_{\rm S}^{(0)\,
2}}\left(\Frac{c_m}{c_d}-1\right)\right] \,
\nonumber
\\
\label{eq.GammaS-RChT}
\\
&=&  \, \Frac{3 \, c_d^{{\rm eff}\, 2} }{16\pi \, f_\pi^4}\, \left[1
\, + \, \Frac{m_\pi^2}{M_{\rm S}^{(0)\,
2}}\left(\Frac{4c_m}{c_d}-6\right) \, +\, \cO(m_\pi^4)\right] \, .
\nonumber
\end{eqnarray}
The global coefficients provides that chiral limit of
$\Gamma_{\rm S}/M_{\rm S}^3$ found in the previous section. The chiral
corrections are there given in terms of the combination of couplings
\begin{equation}
(6+ \alpha_S) \, =\,
6\, +\, \left[ \Frac{M_S^5}{\Gamma_S}\,
\Frac{d}{dm_\pi^2}\left(\Frac{\Gamma_S}{M_S^3}\right)\right]_{m_\pi^2=0}
%
%
\,=\, 2 \, \delta c_d \, -\, 4\, \delta f\, + \, \Frac{f^2}{c_d^2}\, \delta f\, .
\label{eq.alphaS-RChT}
\end{equation}

It is possible to carry a similar analysis for the vector meson,
expressing the $V^r\to \pi^r\pi^r$ in terms of the effective
parameter $G_V^{\rm eff}$. Though the explicit form of $G_V^{\rm
eff}=G_V\left[1+\delta G_V\frac{m_\pi^2}{M_{\rm V}^{(0)\, 2}}\right]$
is not given in this paper, we can write:
\begin{eqnarray}
\Frac{\Gamma_{\rm V}}{M_{\rm V}^3}\, &=& \, \Frac{ G_V^{{\rm eff}\, 2}\,
\rho(M_{\rm V}^2)^3}{48 \pi f_\pi^4}\, \nonumber
\\
\nonumber\\
&=&\, \Frac{G_V^{{\rm eff} \, 2}}{48\pi f_\pi^4}\, \left[ 1\, -  \,
\frac{6\, m_\pi^2}{M_{\rm V}^{(0)\, 2}}\, \,+\, \, \cO(m_\pi^4)
\right]\, ,
\label{eq.GammaV-RChT}
\end{eqnarray}
which gives $(6+\alpha_V)=2 \, \delta G_V- 4\, \delta f
\frac{M_V^{(0)\,2}}{M_S^{(0)\,2}}$.

Gathering all the information from R$\chi$T in
Eqs.~(\ref{eq.GammaS-RChT})--(\ref{eq.GammaV-RChT}), one gets for the modified-KSRF
relation in Eq.~(\ref{sumksrf}) and the new $\alpha_V-\alpha_S$ relation
in Eq.~(\ref{sumalpha}) the result
\begin{eqnarray}
&\Frac{3 \, G_V^2}{f^4}\, +\, \Frac{2\, c_d^2}{f^4}  \, =\, \Frac{1}{f^2}\, ,  &
\label{eq.RChT-rel1}
\\
\nonumber
\\
&\Frac{3 \, G_V^2}{f^4}\, \left[\Frac{2 \, \delta
G_V}{M_V^{(0)\,\,2}} \, - \,\Frac{ 4\, \delta f}{M_S^{(0)\,\, 2}}
\right] \, +\, \left[ \Frac{2 c_d^2}{f^4}\, \Frac{ (2 \delta c_d -4
\delta f)}{M_S^{(0)\,\, 2}} \, +\, \Frac{1}{f^2}\, \Frac{2\, \delta
f}{M_S^{(0)\,\, 2}}\right] = 0\, ,\nonumber \\ \label{eq.RChT-rel2}
&
\end{eqnarray}
where a global factor $1/16\pi$ has been simplified with respect to Eqs.~(\ref{sumksrf})
and (\ref{sumalpha}).
It is not difficult to put the two former equations together into the single  relation
\begin{equation}
\Frac{ 3 \, G_V^{{\rm eff}\, 2}}{f_\pi^4}\, \, + \, \, \Frac{2 \,
c_d^{{\rm eff}\, 2}}{f_\pi^4}\, \,=\,  \, \Frac{1}{f_\pi^2} \,  \, .
\label{eq.RChT-constraint}
\end{equation}
The leading order in its $m_\pi^2$ expansion provides
Eq.(\ref{eq.RChT-rel1})
and its $\cO(m_\pi^2)$ term produces Eq.~(\ref{eq.RChT-rel2}).
A last simplification of a
global factor $1/f_\pi^2$ is left for the reader.
It is remarkable that both resonance constraints are actually governed in R$\chi$T
by the ratios $\frac{c_d^{{\rm eff} \, 2}}{f_\pi^2}$ and
$\frac{G_V^{{\rm eff} \, 2}}{f_\pi^2}$.

Once again, the analysis of the
$\pi\pi$--scattering amplitude at high energies
allows a better understanding of our sum-rule result. We find,
\begin{equation}
T(s)^1_1\, = \,
\Frac{s}{96 \pi f_\pi^2}\,
\left[ 1\, -\,  \Frac{3 \, G_V^{{\rm eff} \, 2}}{f_\pi^2}
\, - \, \frac{2\, c_d^{{\rm eff}\, 2}}{f_\pi^2}\right]\,\,+\,\,\, \cO(s^0)\, .
\end{equation}
Identical results are found for the $IJ=00$ and $IJ=20$ channels.

%
%
%
%

\section{The scalar resonance at $N_C=3$ and $N_C\to \infty$}\label{scalarphys.}

Historically, the understanding on the scalar sector is much less
clear than the vector sector. In Ref.~\cite{XZ00} it is demonstrated
that (when $N_C=3$) a light and broad scalar resonance (the $\sigma$
meson) dominates at low energies in the IJ=00 channel and takes an
essential role to adjust chiral perturbation theory to experiments.
The pole location is estimated in~\cite{zhou} using the dispersion
representation Eq.~(\ref{param}), which are in good agreement with
the more rigorous Roy equation analysis~\cite{CCL06}. Under this
situation it is worthwhile to investigate the role of these light
and broad resonances.

It is not clear, however, what is the nature of this $\sigma$ meson
and different opinions exist on its large $N_c$ behavior. The
$\sigma$ meson may even be considered as a dynamically generated
resonance and decouples in some way from low energy physics when
$N_C$ is large~\cite{pelaez06}. For example, the $K$ matrix unitarization of the
current algebra term yields a $\sigma$ pole in the chiral limit with
the following pole location:
 \be \label{ca} z_\sigma \simeq 16i\pi
f_\pi^2\ .
 \ee
 This `current algebra  $\sigma$' maintains an unusual property:
 It flies away on the complex $s$ plane meanwhile
 it contributes to
 the $r.h.s.$ of Eq.~(\ref{sumksrf}) in the large $N_C$
 limit. Nonetheless,
 such a pole does not contribute to the sum rule for $L_3$, i.e., Eq.~(\ref{sumL3}).
 Indeed  the existence of poles which moves
  to $\infty$ can not be
 excluded using pure  $N_c$ counting rule. However in
the $s$ channel such a pole contributes, in the chiral limit, a term
 \be\label{exotic}
  T^{\rm sR}(s)=\frac{{s\over 16\pi f^2}}{1-i\rho\frac{s}{16\pi
  f^2}}
   \ee
to the $r.h.s.$ of Eq.~(\ref{sumTsR}), according to
Eqs.~(\ref{resp}) and (\ref{MandGamma}).  However, unlike the
ordinary  narrow resonances,  crossing symmetry is not fulfilled.
Beside this, the `current algebra $\sigma$' in Eq.~(\ref{exotic})
contributes $1/16\pi f^2$ to the r.h.s. of Eq.~(\ref{sumksrf}). This
is  misleading since the KSRF relation Eq.~(\ref{sumksrf}) tells
where the factor $1/16\pi f^2$ comes from. Furthermore, the
unitarization of the current algebra amplitude produces unphysical
poles $z_\rho=96 i \pi f^2$ in the second Riemann sheet and
$z_{(20)}=32 i\pi f^2$ in the first Riemann sheet. This leads to an
incorrect interpretation of the KSRF relation.

It is important to notice that the behavior of $\sigma$--meson  must be
totally different in the case when $N_c\to \infty$. It is noticed that
the $N_c$ dependent pole trajectory for $\sigma$ behaves very
differently from that of $\rho$~\cite{pelaez06}. This phenomenon is
re-investigated in Ref.~\cite{sun05}. It is found that, even
though the $\sigma$ pole trajectory is bent from the expected large--$N_C$
behaviour, it can finally fall down to the real axis at $N_C\to \infty$ and,
hence, be relevant at large--$N_C$.
It is argued in Ref.~\cite{guo06} that the bent
structure of the $\sigma$ pole trajectory itself is not sufficient
to demonstrate that the $\sigma$ pole is dynamically generated.
Although these investigations are based on models and other assumptions, they
show that this alternative scenario should not be ruled out.

We want to finish with a numerical analysis of
Eqs.~(\ref{sumksrf}),~(\ref{sumL2})~and~(\ref{sumL3}), \ where \ we \ will \ consider \
the inputs  $f=88$~MeV, $M_\rho=770$~MeV, $\Gamma_\rho=146  $~MeV. Since we
assume that the scalar becomes a narrow--width state at $N_C\to\infty$,
the values of $M_\sigma$ and $\Gamma_\sigma$ should be different
from their corresponding values at $N_C=3$. Here we adopt a rather
exaggeratory value of the scalar parameters, $M_\sigma=700$~MeV and
$\Gamma_\sigma=500$~MeV.
For the r.h.s. of the modified--KSRF relation in Eq.(\ref{sumksrf}),
one has  (in units of GeV$^{-2}$)
\begin{equation}
\Frac{9 \, \Gamma_V^{(0)} }{M_V^{(0)\,\, 3}  } \, +\,
\Frac{2 \, \Gamma_S^{(0)} }{3\, M_S^{(0)\,\, 3}  } \,
\, \simeq \,2.9 \, +\, 1.0 \, \,,
\end{equation}
where the first term on the right--hand
side comes from the vector contribution and the second one from the
scalar. From the modified--KSRF relation, one would expect their sum to be equal to
${1}/{16 \pi f^2}\simeq 2.6$~GeV$^{-2}$.
Although these large--$N_C$ estimates are rough, they
suggest that there is almost no room for the scalar contribution to
the $r.h.s$ of Eq.~(\ref{sumksrf}).
Thus, in the picture suggested
in Ref.~\cite{XZZ06}, the bare $\sigma$ mass turns out to be of the order of
$M_\sigma \sim 1$~GeV, resulting the scalar contribution indeed suppressed by
the large mass and becoming the modified--KSRF relation
insensitive to the value of $\Gamma_\sigma$.

Our numerical prediction for $L_2$ and $L_3$ at large--$N_C$  is
\begin{eqnarray}
10^3\, \cdot \,L_2 &\simeq& \,1.2 \, \, ,
\label{eq.L2num}
\\
10^3\, \cdot\, L_3 &\simeq&\,  -3.7\, +\,  1.5 \,\, ,
\label{eq.L3numn}
\end{eqnarray}
where the first contribution to $L_3$ comes from
the vector meson and the last one from the scalar.
This can be compared to the one--loop experimental determination,
$10^3\, L_2^r=1.35\pm 0.3$, $10^3\, L_3^r=-3.5\pm 1.1$~\cite{fitoneloop},
and to Bijnens' $\cO(p^6)$ result,
$10^3\, L_2^r=0.73\pm 0.12$, $10^3\, L_3^r=-2.35\pm 0.37$~\cite{fit10}.
It is possible to isolate the scalar resonance contribution to the LECs by considering
an appropriate combination of Eqs.~(\ref{sumL2})~and~(\ref{sumL3}):
\begin{equation}
L_3\, +\, 3L_2\,\, =\,\, \Frac{8\pi  f^4\, \Gamma_\sigma}{3 M_\sigma^5}
\,\, > \,\, 0\,  \,,
\end{equation}
which, for our input values $M_\sigma=700$~MeV, $\Gamma_\sigma=500$~MeV, yields
\begin{equation}
L_3\, +\, 3L_2\,\, \simeq \,\, 1.5\,\cdot \, 10^{-3}\, \, .
\end{equation}
The experimental determinations for $L_2$ and $L_3$ in $\chi$PT
provide the upper bound
$L_3^r+3 L_2^r \lsim  1.9 \cdot 10^{-3}$ at one loop~\cite{fitoneloop} and
$L_3^r+3 L_2^r \lsim  0.36 \cdot 10^{-3}$ at $\cO(p^6)$~\cite{fit10}.
This indicates that, at large--$N_C$, either $\Gamma_\sigma$ is small or
$M_\sigma$ becomes large.
For example, for $\Gamma_\sigma=500$~MeV,
the smallest value for the mass is $M_\sigma\simeq 670$~MeV
if the one-loop upper bound is assumed, and  ${ M_\sigma\simeq 930}$~MeV if
we take the $\cO(p^6)$ result.
Nevertheless, it is important to recall
that experimental determinations of the LECs differ from the corresponding values
at large--$N_C$ due to subleading corrections in $1/N_C$~\cite{nloNC},
so one should be cautious about these bounds.

In any case, the safe conclusion from Eq.~(\ref{sumksrf}) is that
the scalar meson takes a numerically minor role in the KSRF relation
when $N_C$ is large. The situation can be quite different in the
$N_C=3$ case. For instance, the present work shows that the $IJ=00$
scattering length is dominated by the crossed-channel $\rho$
exchange at large--$N_C$. However, the phenomenological analysis of
the $IJ=00$ experimental data is found to be  dominated by the
$s$--channel scalar contribution~\cite{zhou}.

\section{Discussions and Conclusions}

In this paper we started from a variation of partial wave dispersion
relation (the PKU form) and demonstrated that it is reduced to the
standard once subtracted partial wave dispersion relation (PWDR) in
the narrow width approximation or in the leading order of $1/N_c$
expansion. Matching the resonance contribution calculated from PWDR
to the low energy chiral amplitudes up to $\cO(p^4)$ leads to a
set of resonance sum rules. They include  the KSRF relation, two
sum rules for the low energy constants $L_2$, $L_3$ and a new
relation between resonance couplings, Eq.~(\ref{sumalpha}).

We made a rather detailed examination of the new relation in various
resonance chiral lagrangians and found that it is not always
trivially fulfilled. Hence it provides a useful novel constraint for
the construction of the hadronic action. The origin of this
constraint is understood: It comes from the requirement of chiral
symmetry and a proper high energy behavior of the scattering
amplitude. We start from an $S$ matrix theory point of view, which
is crucial to provide a rigorous and systematic way to derive the
sum rules, independently of the realization of the resonance
lagrangian. Our investigation provides a clearer understanding to
the KSRF relation and generalizes it beyond the leading chiral
order. We also discussed the $N_c$ property of the $\sigma$ meson
and conclude that, unlike the case when $N_c=3$, it takes a
numerically negligible role when $N_c\to \infty$.

 \vspace{1cm}

{\bf\large Acknowledgements}:
  This work is support in part by National
 Nature Science Foundations of China under contract number
 10575002,
  10421503. 
We wish to acknowledge the useful comments from L.-Y.~Xiao and J.~Portol\'es.


\end{document}